\documentclass[10pt,conference]{IEEEtran} 
\IEEEoverridecommandlockouts

\usepackage[utf8]{inputenc}
\usepackage[hidelinks]{hyperref}
\usepackage{xcolor}
\usepackage{amsmath, amssymb, amsfonts}
\usepackage{amsthm}
\usepackage{times}
\usepackage{soul}
\usepackage{graphicx}
\usepackage{subcaption}
\usepackage{float}
\usepackage{booktabs} 
\usepackage{threeparttable}
\usepackage{tabularx}
\usepackage{multirow}
\usepackage{siunitx} 

\usepackage{algorithm}
\usepackage{algorithmic}

\usepackage{cite}

\usepackage[most]{tcolorbox}
\newtcolorbox{myresultbox}[1][]{
    colback=gray!20,
    colframe=black!75,
    fonttitle=\bfseries,
    title=Result,
    #1
}
\renewcommand{\thetable}{\arabic{table}}

\usepackage[margin=0.75in]{geometry}
\usepackage{cite}

\def\BibTeX{{\rm B\kern-.05em{\sc i\kern-.025em b}\kern-.08em
    T\kern-.1667em\lower.7ex\hbox{E}\kern-.125emX}}
\begin{document}
\renewcommand{\arraystretch}{1.5}
\setlength{\tabcolsep}{10pt}
\title{\LARGE \bf Line-level Semantic Structure Learning for Code Vulnerability Detection}

\author{
    \IEEEauthorblockN{
        Ziliang Wang\IEEEauthorrefmark{1}, 
  Ge Li\IEEEauthorrefmark{1}\thanks{Corresponding Author: Ge Li (lige@pku.edu.cn)},
        Jia Li\IEEEauthorrefmark{1}, 
        Yihong Dong\IEEEauthorrefmark{1}, 
        Yingfei Xiong\IEEEauthorrefmark{1}, 
        Zhi Jin\IEEEauthorrefmark{1}
    }
    \IEEEauthorblockA{\IEEEauthorrefmark{1}Key Lab of High Confidence Software Technology, MoE, School
of Computer Science, Peking University, Beijing, China \\
    Email: \{wangziliang, lige,xiongyf, zhijin\}@pku.edu.cn}
    \{lijia, dongyh\}@stu.pku.edu.cn
}

\maketitle

\begin{abstract}
Unlike the flow structure of natural languages, programming languages have an inherent rigidity in structure and grammar.
However, existing detection methods based on pre-trained models typically treat code as a natural language sequence, ignoring its unique structural information. This hinders the models from understanding the code's semantic and structual information.
To address this problem, we introduce the Code Structure-Aware Network through Line-level Semantic Learning (CSLS), which comprises four components: code preprocessing, global semantic  awareness, line semantic awareness, and line semantic structure awareness.
The preprocessing step transforms the code into two types of text: global code text and line-level code text.
Unlike typical preprocessing methods, CSLS retains structural elements such as newlines and indent characters to enhance the model's perception of code lines during global semantic  awareness.
For line semantics structure awareness, the CSLS network emphasizes capturing structural relationships between line semantics.
Different from the structural modeling methods based on code blocks (control flow graphs) or tokens, CSLS uses line semantics as the minimum structural unit to learn nonlinear structural relationships, thereby improving the accuracy of code vulnerability detection.
We conducted extensive experiments on vulnerability detection datasets from real projects. 
The CSLS model outperforms the state-of-the-art baselines in code vulnerability detection, achieving 70.57\% accuracy on the Devign dataset and a 49.59\% F1 score on the Reveal dataset. At the same time, the proposed preprocessing method significantly improves the performance of the existing baseline model, which provides a new revelation for the follow-up research in this direction.
These results demonstrate the importance of preserving and utilizing code structure information to improve the performance of code vulnerability detection models.
\end{abstract}


\section{Introduction}
Software code vulnerabilities refer to weaknesses in code that can be exploited, leading to severe consequences such as unauthorized information disclosure and cyber extortion~\cite{fu2022linevul, thapa2022transformer}. 
The growing magnitude of this issue is highlighted by recent statistics: in the first quarter of 2022, the US National Vulnerability Database (NVD) reported 8,051 vulnerabilities, a 25\% increase from the previous year~\cite{cheng2022path}. 
Additionally, a study found that 81\% of 2,409 analyzed codebases contained at least one known open-source vulnerability~\cite{Synopsys}.
The widespread nature and increasing number of these vulnerabilities underscore the urgent need for robust, automated vulnerability detection mechanisms.
Implementing such systems is crucial for enhancing software security and preventing a range of potential threats~\cite{fu2022linevul, thapa2022transformer, jang2014survey, johnson2011guide}.

Existing literature on vulnerability detection models can be broadly categorized into two main types: (1) traditional detection models~\cite{yamaguchi2015pattern,yamaguchi2017pattern}, and (2) deep learning (DL)-based models~\cite{duan2019vulsniper,russell2018automated,li2018vuldeepecker,dam2017automatic}. 
Traditional detection models often require experts to manually develop detection rules~\cite{Checkmarx,Flawfinder}.
This approach is labor-intensive and struggles to maintain low rates of false positives and false negatives~\cite{li2018vuldeepecker,li2021sysevr}.
In contrast, deep learning (DL)-based detection methods learn vulnerability patterns from training datasets~\cite{li2018vuldeepecker,dam2017automatic,lin2017poster,steenhoek2024dataflow}, eliminating the need for manual heuristics and enabling autonomous feature identification.
They avoid manual heuristic methods and autonomously learn and identify vulnerability features.

\begin{figure}
\centering
\includegraphics[width=0.48\textwidth]{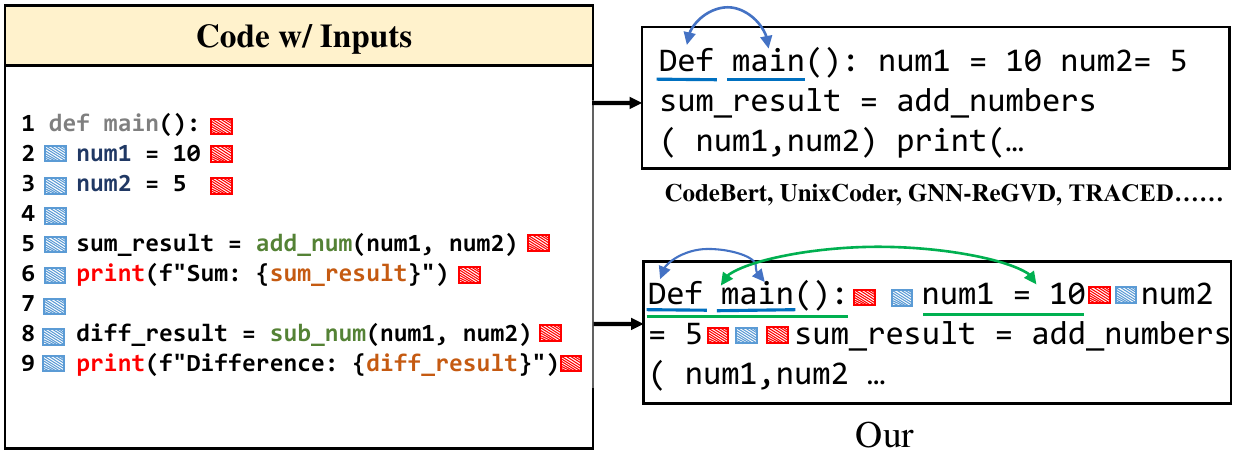}
\caption{Existing preprocessing methods make the model focus only on the positional relationship between tokens, making it difficult to identify the structural relationship between code lines. CSLS facilitates the model in identifying code lines and learning relationships between line semantics.}
\label{fig:old}
\vspace{-1em}
\end{figure}
To further enhance the understanding of vulnerability semantics, recent studies have incorporated code structure into pre-trained code models~\cite{nguyen2022regvd,ding2024traced,zhang2023vulnerability}.
The Devign~\cite{zhou2019devign} method combines the Abstract Syntax Tree (AST), Control Flow Graph (CFG), Data Flow Graph (DFG), and code sequence and other structures to introduce structural information. 
The ReGvd \cite{nguyen2022regvd} is a more elaborate way of modeling the structure of code based on the GraphCodeBert~\cite{guo2020graphcodebert}, compared to modeling the structure based on code blocks (multiple lines of code) by building a graph with tokens as the smallest unit.
In the latest research, TRACED~\cite{ding2024traced} takes the execution path information into account in the pre-trained model, thus providing an effective vulnerability detection model.

However, these methods usually preprocess the code into natural language text and remove structural elements, as shown in Figure 1~\cite{feng2020codebert,guo2022unixcoder}\footnote{https://github.com/microsoft/CodeXGLUE}~\cite{steenhoek2024dataflow}\footnote{https://github.com/daiquocnguyen/GNN-ReGVD}~\cite{ding2024traced}\footnote{https://github.com/ARiSE-Lab/TRACED\_ICSE\_24.git}.
The loss of these structuring elements often leads to models that focus only on the relationships between tokens and ignore the benefit of line structure to the model's understanding ability.
For example, a newline character can tell the model that the statement is over, and a blank line indicates that the code above and below it may have different functions, such as variable definition and function implementation.
Although these methods reintroduce code structure information in various ways, whether based on code blocks or tokens, it is still challenging to comprehensively construct the structural information of complex code texts using extensive modeling rules (such as CFG)~\cite{wu2022survey,phan2017convolutional}. 
For example, two completely unrelated control flow graph (CFG) code blocks can also lead to vulnerabilities due to resource contention (e.g., race conditions)~\cite{carr2001race}.

\emph{\textbf{Our Approach.}}
In this paper, we explore the extension of line-level code text to semantic space with the help of pre-trained models, and take line semantics as nodes to model the complex nonlinear relationship between code lines, so as to develop an effective vulnerability detection model.
Therefore, in \emph{code preprocessing}, we will preserve the structural elements in the code text and split the code into a series of lines.
Then we propose \emph{CSLS model} to model the structure of line semantics in high-dimensional semantic space.

\emph{Code Pre-processing Process.} To address the issues mentioned above, we proposed a new code preprocessing method to enhance code structure perception in vulnerability detection tasks. This process results in two different code texts:
(1) \emph{Global Code Text.} We replace the previous preprocessing with a simpler method to obtain global word segmentation, as shown in Figure 1. CSLS preserves line breaks and all whitespace before tokenization.
(2) \emph{Line-level Code Text.} By preserving code structure information, CSLS splits the code snippets into lines and generates a line-level token array for each snippet.
This approach preserves the original code structure while enabling the model to learn from both global and line-level contexts.

\emph{CSLS Model.} To further enhance code structure perception, we propose a new vulnerability detection network architecture using code models to learn code semantics and structure from different code texts.
(1) \emph{Global Semantic-Aware.} The model takes the global code text as input to complete the modeling of global semantic and structural information. Benefiting from the preservation of structural elements, further understanding of the structure enhances the performance of global semantic understanding(see V.C).
(2) \emph{Line Semantic-Aware.} A pre-trained code model uses line-level code as input to capture the semantics of each line.
(3) \emph{Line Semantic Structure-Aware.} A Transformer module models the line semantics of a code fragment, using line numbers as positional encoding. Its multi-head attention mechanism learns nonlinear structural relationships between line semantics.
Finally, the CSLS model analyzes code vulnerability risks from three perspectives: global semantics, line semantics, and line semantic structures, achieving high-precision vulnerability detection.

\emph{\textbf{Results. }}  
%
With our \emph{Code Pre-processing Process}, we report state-of-the-art results on popular or representative code models: CodeBERT's~\cite{feng2020codebert} accuracy increased from 63\% to 65\% on the Devign dataset\cite{zhou2019devign}, and UniXcoder-base's accuracy rose from 64.8\%\cite{nguyen2022regvd} to 68.8\%. Other datasets, such as ReVeal\cite{chakraborty2021deep}, also show positive effects.
Our proposed CSLS further enhances vulnerability detection accuracy. We achieved state-of-the-art results with CSLS, reporting 70.57\% accuracy on the Devign dataset, and 91.86\% accuracy and a 49.59\% F1-score on the ReVeal dataset.
In summary, the main contributions of this paper are:
\begin{itemize}
\item [a)] We propose an enhanced code preprocessing method for vulnerability detection tasks, highlighting the flaws of existing preprocessing methods and emphasizing the importance of preserving code structure information.

\item [b)] We propose a vulnerability detection network architecture that enhances code structure awareness. The model uses the pre-trained model to learn the line-level semantics, uses the transformer to learn the nonlinear structural relationship of the line-level semantics, and combines it with the global information to achieve high-precision vulnerability detection.

\item [c)] We evaluate the proposed preprocessing method on two real datasets. The experimental results show that preserving code structure information effectively improves the performance of existing methods and provides new benchmark data for vulnerability detection.

\item [d)] We evaluate our network architecture on two real-world datasets. The results show that the proposed CSLS architecture further improves the accuracy of code vulnerability detection by perceiving code line-level semantic structures
\footnote{Our anonymous  replication package (data and code) :https://figshare.com/articles/dataset/CSLS\_model\_code\_and\_data/26391658}.
\end{itemize}

\section{RATIONALE}
In this section, we discuss the background of vulnerability detection using code structure information. This includes explaining the core motivation of our approach in terms of both the object being modeled and the modeling approach.
We also compare related works that incorporate code structure in model fine-tuning to highlight the novelty of our work.

\subsection{Control Flow Graph (CFG) based on Code Block}

A Control Flow Graph (CFG) illustrates the control flow within a program, where each node represents a basic block, and edges denote the control flow between these blocks. A basic block is a sequence of statements or instructions with a single entry and exit point, typically consisting of multiple lines of code executed sequentially.
\emph{(1) Basic Block Identification:}
The first step is to identify the basic blocks, which are determined by control flow instructions like conditional and unconditional jumps. A basic block usually consists of multiple lines of code executed sequentially between a single entry and exit point.
\emph{(2) Construction of Nodes and Edges:}
Each basic block is represented as a node in the CFG, with edges added based on control flow instructions.
\emph{(3) Marking Entry and Exit Nodes: }
Identify and mark the program's entry and exit nodes, with the entry node indicating the program's start and the exit node its end.
Devign~\cite{zhou2019devign}, a popular vulnerability detection method, uses the Abstract Syntax Tree (AST), Control Flow Graph (CFG), and Data Flow Graph (DFG) of code snippets as inputs, combining these representations for vulnerability detection.

\subsection{Undirected Graphs based on Code Tokens}
An existing fine-tuning method (ReGVD)~\cite{nguyen2022regvd} incorporates code structure into the process, primarily using graph neural networks (GNNs) to detect vulnerabilities in source code by representing the code structure as a graph.
Graphs are constructed in two main ways:
\emph{(1) Unique Token-Focused Graph: } Nodes represent unique tokens in the source code.
\emph{(2) Index-Focused Graph: }Each position (index) in the token sequence of the source code is represented as a node, and edges are created between positions that co-occur within a fixed-size sliding window.

ReGVD uses a more flexible and general approach to building graphs based on token co-occurrence to capture semantic and structural relationships, differing from CFGs.





\subsection{The Novelty of Our Work based on Line Semantics}
These methods focus on the role of code structure in vulnerability detection from various perspectives~\cite{zhou2019devign,nguyen2022regvd,guo2020graphcodebert}. 
However, heuristic approaches to structural modeling in low-dimensional spaces are limited. 
The structure of vulnerabilities caused by complex interactions between multiple lines of code is usually difficult to model by conventional structure modeling methods.
Inspired by the state selection network~\cite{gu2023mamba}, CSLS encodes lines of code into a high-dimensional space, modeling nonlinear structures in the semantic space with line semantics as the basic unit.
Compared to existing structure-aware vulnerability detection methods, CSLS offers the following advantages:
By preserving structural elements during global semantic modeling, CSLS enhances the model's ability to recognize code context and distinguish code blocks. 
Furthermore, in the line semantic structure-aware process, we model complex nonlinear relationships between lines of code in a high-dimensional semantic space. 
This dual approach allows our method to capture intricate structural relationships within the code, thereby offering a more comprehensive and precise understanding of the code structure.

\section{Approach} In this section, we introduce the Code Structure-aware Network through Line-level Semantic Learning (CSLS).
Figure 2 provides an overview of CSLS, which comprises four main phases: (1) Data Preprocessing, (2) Line-level Semantic-aware, (3) Line semantic Structure-aware, and (4) Global Semantic-aware.
\begin{figure*}
\centering
\includegraphics[width=0.98\textwidth]{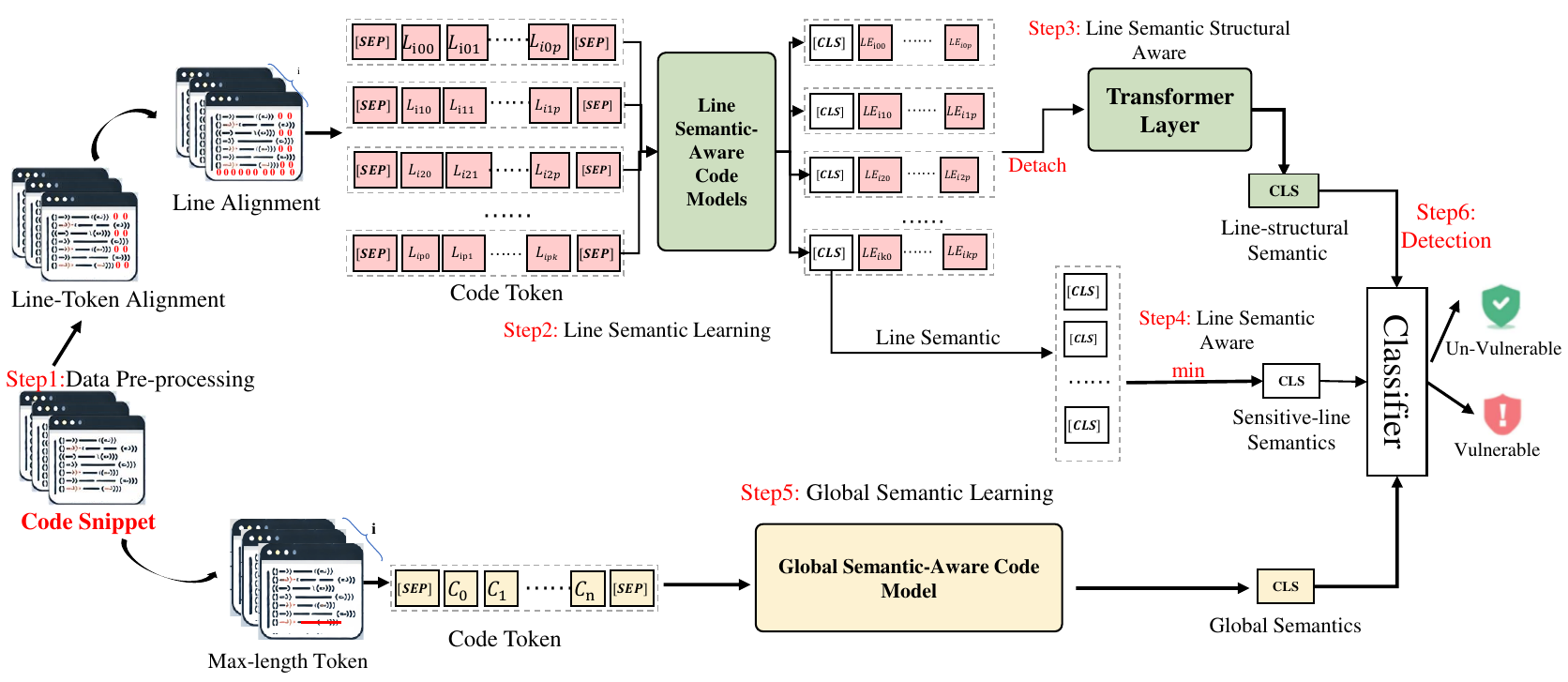}
\caption{The CSLS framework implements vulnerability detection by capturing three different semantics: (1) Line-level Structural Semantics (Step 3), (2) Sensitive-Line Semantics (Step 4), and (3) Global Semantics (Step 5).
}
\label{fig:main}
\vspace{-1em}
\end{figure*}

\subsection{Data Preprocessing}

We denote the historical vulnerability dataset as \( L = (P, Y) \), where \( p_i \) represents a code snippet, and \( y_i \) represents the corresponding binary label (i.e., vulnerable or secure) for the \(i\)-th snippet, with \( 0 < i \leq M \). Here, \( M \) is the total number of code snippets. The labels \( y_i \) can take values of either 0 or 1, where \( y_i = 0 \) indicates that the code is free of vulnerabilities, and \( y_i = 1 \) indicates the presence of a vulnerability in the code.

\textbf{Global Preprocessing.} Unlike existing methods that remove structural elements, we only truncate code fragments to fit the input size constraints of the model. This step is formalized as follows: 

\begin{equation}
C_{i} = C_{i_0} C_{i_1} \ldots C_{i_n}, \quad 0 < i \leq M
\end{equation}

Here, \( C_{i_n} \) represents the \( n \)-th token of the \( i \)-th code fragment, with \( n \) being the maximum input length allowed by the code model.

\textbf{Code Line Preprocessing.} 
The number of lines in a code fragment typically exceeds the number of snippets. 
We define \( L_{i_n} \) as the \( n \)-th line of the \( i \)-th code fragment.
To expedite the learning process for the code model, we implement the following preprocessing steps for each training batches:

%
\emph{Line Alignment:} 
For all snippets in the same training batch, we align the number of lines in each snippet. 
%
Statistical analysis shows that the average number of lines of code in the Devign dataset training set is 62.21 under the 1024 length limit.
Therefore, CSLS presets \( MAX(k) = 100 \) lines per code fragment split. For all code fragments \( n \) in the same batch, we use \( k = \max(\text{len}(n.\text{line}), 100) \).

\begin{equation}
L_{i} = L_{i_0} L_{i_1} \ldots L_{i_k}, \quad 0 < k \leq 100
\end{equation}

Here, \( L_{i_k}  \) represents the \( k \)-th line of the \( i \)-th code fragment, with \( k \) being the maximum number of line.

%
\emph{Line-Token Alignment:} Each line in the code snippet is tokenized individually. Tokens are then padded to ensure uniform length. For snippets within the same batch, we align the number of tokens per line.
Statistical analysis shows that p=20 can cover 93.41\% of the lines of code under the 1024 input limit.
Thus, CSLS presets the maximum number of tokens per line to \( p = 20 \). 

\begin{equation}
L_{i_k} = T_{i_{k0}} T_{i_{k1}} \ldots T_{i_{kp}}, \quad p=20
\end{equation}

Here, \( L_{i_{kp}} \) represents the \( p \)-th token of the \( k \)-th line of the \( i \)-th code fragment, with \( p \) being the maximum number of tokens per line.

The preprocessed snippets have the same number of lines of code and number of tokens per line.
This approach allows the model to generate semantics for all lines of the same training batch at the same time instead of learning one line at a time, which will greatly improve the training speed and reduce the learning cost.

\subsection{Line-level semantic and structure aware}
\textbf{Line Semantic Learning.}
After the code preprocessing, we obtain an array of tokens based on line segmentation for all code fragments. 
And, in the same batch,  this array has a consistent shape of $[b,k,p]$. Where $b$ denotes the number of snippets in the batch, $k$ is the number of lines in each snippet, and $p$ is the maximum number of tokens per line.

We flatten the preprocessed token array to form a two-dimensional array $z$ with size $[b*k, p]$. 
This allows us to process an entire batch of data in a single operation without iterating over each line. CSLS uses a line semantic-aware model, and this paper uses the UniXcoder-nine model by default. 
We process $z$ to quickly obtain deep semantic representation through a single batch. The formalization process is as follows:

\begin{equation}
LE=f_l(z,z_{mask}),
\end{equation}
where \( LE \) has the shape \([b \times k, h]\), with \( h = 768 \). Here, \( b \) represents the batch size, \( l \) represents the sequence length, and \( h \) represents the dimension of the hidden layer.
The $f_l$ stands for line semantic-aware model. $z_{mask}$ is provided by the completion process during data pre-processing. The output $LE$ of the line semantic-aware model contains multiple hidden states, and we use the hidden state of the first token of the last layer (usually the [CLS] token) to represent the semantics of each lines.

Finally, to restore the processed data to its original batch structure, we reshape $LE$ to $[b, k, h]$, where each element now contains an embedded representation of the corresponding line of code. This refactoring facilitates subsequent steps such as further analysis or specific line-level tasks.

\textbf{Line-Semantic Structure Aware.} After the above process, CSLS obtains the semantic array $LE$ for each line of the code fragment. In this module, CSLS uses a Transformer model to model line semantics to learn nonlinear structural information between lines.
\begin{figure}[h]
\centering
\includegraphics[width=0.48\textwidth]{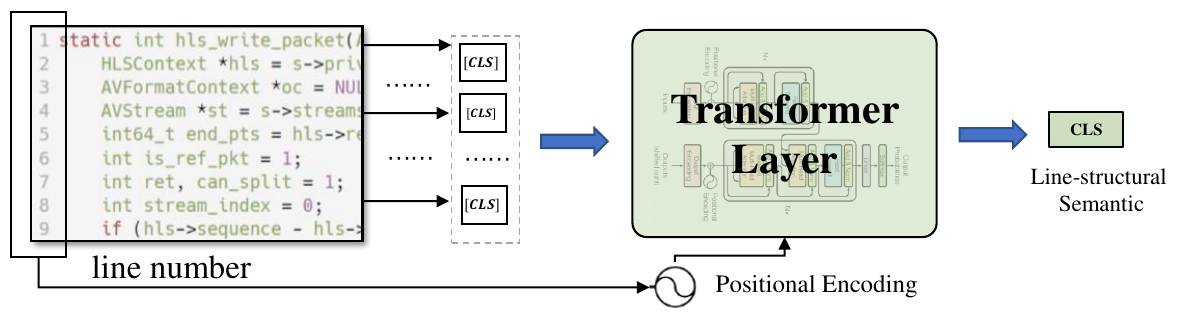}
\caption{A Transformer model is used to capture the semantic  structural relationships between lines.}
\label{fig
}
\vspace{-1em}
\end{figure}

As shown in Figure 3, the input line semantics $LE$ preserves the order of line numbers, and the default positional encoding of the Transformer layer corresponds to these line numbers. 
This alignment allows the Transformer to capture the structural relationships between lines of code. The output of the Transformer layer is then used to assess the vulnerability of the code based on its structural representation. The formalization is as follows:

\begin{equation}
S_{repr}=f_t(D(LE_1..LE_k))
\end{equation}

Where $f_t$ represents a standard Transformer model with 8 layers and 8 attention heads.  
The multi-head attention mechanism within the Transformer enhances this process by allowing each attention head to focus on different aspects of the relationships between line semantics, enabling a more comprehensive and nonlinear modeling of the line-level semantic structure. 

\emph{Semantic Detach.} In this step, we introduce a crucial step  $D(LE)$ by detaching the sentence embeddings from the global encoder before passing them to the Transformer encoder.  
This detachment prevents gradients from flowing back into the global encoder during the optimization of the Transformer encoder.
The purpose of this separation is to ensure that the Transformer encoder focuses exclusively on learning the structural relationships between code lines, rather than reprocessing semantic information already captured by the line-semantic model. 
This delineation of responsibilities allows the line-semantic model to specialize in semantic understanding and the Transformer encoder to specialize in structural dependency modeling. 
%

\textbf{Sensitive-line Semantic Awareness.} In the line semantic-aware module, we obtain the overall semantic representation of each line of code by calculating the average of the 768-dimensional vectors in the semantic representation of each line.
For the code snippets in each batch, we identify the most representative lines of code by computing the minimum of the semantic representation of each line. The specific process is as follows:

For the semantic representation matrix $LE$ for each sentence, we first compute the average of each token over 768 dimensions:

\begin{equation}
LE_{mean} = \frac{1}{h} \sum_{j=1}^{h} LE_{i,j}
\end{equation}
Then, we find the row index with the smallest mean for each batch:
\begin{equation}
L_{min} = Index(\arg \min(LE_{mean}))
\end{equation}

Finally, we select the semantic representation of the row with the smallest mean value from each batch to form the final sensitive-line semantic representation:
\begin{equation}
L_{repr} = LE[L_{min}]
\end{equation}

where $L_{repr}$ represents the semantic representation of a code fragment whose line semantics is closest to 0.
We choose the Top1 way to update the row semantics because the row-level labels are difficult to obtain in the common scenario. In the most extreme case, only one line is vulnerable. So a piece of code is vulnerable if the line with the highest risk (closest to 0) is vulnerable.
We cannot use the each line semantics and fragments labels for loss calculation, because not every line is vulnerable, even for vulnerable code fragments.

\subsection{Global Semantic Awareness}
\textbf{Global Semantic Learning.}
In the global semantic learning module, we use a code model to capture the global semantic information of code fragments. 
We feed the code tokens $C_i$ into the global semantic-aware model (UniXcoder-nine):
\begin{equation}
G_{repr} = f_g(C_i)
\end{equation}

where $G_{repr}$ is the first token CLS of the model output, which usually represents the global semantics.

\textbf{Vulnerability Prediction.} In the vulnerability prediction module, we concatenate the global semantic
$G_{repr}$, the line-sensitive semantic $L_{repr}$, and the line semantic structure representation 
$S_{repr}$ to form the final representation vector $H$:
\begin{equation}
H = [S_{repr}, L_{repr}, G_{repr}]
\end{equation}

Then, we input the final representation vector into a Multi-Layer Perceptron (MLP) classifier for vulnerability prediction:
\begin{equation}
\hat{y} = \sigma(f_m(H))
\end{equation}
Where $\sigma$ is the Sigmoid activation function, and $P$ is the predicted vulnerability probability. The $f_m$ is a classification network.

\subsection{Loss Function}
The loss function adopted for the code models training is the cross-entropy loss~\cite{zhou2019devign}, commonly used in classification problems for its effectiveness in penalizing the predicted labels and the actual labels:
\begin{equation}
H(y, \hat{y}) = -y \log(\hat{y}) - (1 - y) \log(1 - \hat{y})
\end{equation}
where $y$ is the actual label,  $\hat{y}$ is the predicted value.

\section{Study Design}

\begin{table*}[t]
\centering
\caption{Comparison results for different models on Devign and Reveal datasets. }
\label{my-label}
\small
\begin{threeparttable}
\begin{tabularx}{\textwidth}{lXXXXXXXX}
\toprule
 & \multicolumn{4}{c}{\textbf{Devign \cite{zhou2019devign}}} & \multicolumn{4}{c}{\textbf{Reveal \cite{chakraborty2021deep}}} \\  \cline{2-5} \cline{6-9}
 Models& \textbf{Acc} & \textbf{Recall} & \textbf{Prec} & \textbf{F1} & \textbf{Acc} & \textbf{Recall} & \textbf{Prec} & \textbf{F1} \\ \midrule
ChatGPT 3.5 COT&	49.83&	32.24&	33.00&	30.61&	63.72&	26.34&	30.54&	27.70\\
ChatGPT 4o COT&	53.73&	7.46&	45.94&	4.06&	20.97&	22.17&	97.33&	12.51\\
Devign & 56.89 & 52.50 & 64.67 & 57.59 & 87.49 & 31.55 & 36.65 & 33.91 \\
ReGVD & 61.89 & 48.20 & 60.74 & 53.75 & 90.63 & 14.47 & 64.70 & 23.65 \\
CodeBERT & 63.59 & 41.99 & 66.37 & 51.43 & 90.41 & 25.87 & 54.62 & 35.11\\
UniXcoder-base & 65.77 & 51.55 & 66.42 & 58.05 & 90.50 &39.91& 53.52& 45.72 \\ 
CodeT5+& 65.62 & 55.29 & 64.73 & 59.64 & 90.94 & 31.14& 59.16& 40.80 \\ 
TRACED & 64.42 & 61.27 & 60.03 & 61.05 & 91.11 & 21.49 & 68.05 & 32.66 \\ 
UniXcoder-nine &66.98 & 56.33 & 66.63 & 61.05  & 90.72 & 33.77 & 56.20 & 42.19\\ 
\midrule
CSLS & 70.57 & 59.36 & 71.70 & 64.95 &  91.86&39.91 &65.46& 49.59\\ \bottomrule
\end{tabularx}
\end{threeparttable}
\vspace{-2em}
\end{table*}

\subsection{Datesets}
To assess the effectiveness of CSLS, we utilize two datasets from real-world projects: (1) Devign ~\cite{zhou2019devign} and (2) Reveal ~\cite{chakraborty2021deep}.

\textbf{Label balanced.} The Devign dataset, sourced from a graph-based code vulnerability detection study ~\cite{zhou2019devign}, comprises function-level C/C++ source code from the well-known open-source projects QEMU and FFmpeg. Following the methodology outlined by Li et al. ~\cite{zhou2019devign}, the dataset is divided into training, validation, and testing sets using a standard 80:10:10 split. Security researchers labeled the vulnerable code through a meticulous two-stage review process. The ratio of positive and negative samples in this dataset is close to 1:1, which is label balanced.

\textbf{Label unbalanced.} The REVEAL dataset, as detailed in ~\cite{chakraborty2021deep}, addresses issues of data redundancy and unbalanced class distributions in existing datasets, making it suitable for software vulnerability detection tasks. This dataset includes source code from the Linux Debian kernel and Chromium projects and features an imbalanced label distribution, with a 10:1 ratio of normal to vulnerable code fragments. The REVEAL dataset also follows an 80:10:10 split for training, validation, and testing.
Throughout the experiments, the proportion of positive and negative samples in the training, validation, and testing sets remained consistent with the original dataset distributions.

\subsection{Performance Metrics}
In the process of evaluating the performance of the model, the proposed method employs four metrics\cite{zhou2019devign}:

\textbf{Precision}: This metric is defined as the quotient of true positives (TP) and the sum of true positives and false positives (FP), providing a measure of the accuracy of instances identified as positive. Formally, it is given by:
$Precision = \frac{TP}{TP + FP}.$

\textbf{Recall}: Recall measures the fraction of actual positive instances that are correctly identified. It is calculated as the ratio of true positives to the sum of true positives and false negatives (FN):
$Recall = \frac{TP}{TP + FN}.$

\textbf{F1 Score}: The F1 is the harmonic mean of precision and recall, providing a single metric that balances both concerns:
$F1 = 2 \times \frac{Precision \times Recall}{Precision + Recall}.$

\textbf{Accuracy}: Accuracy reflects the proportion of true positive and true negative instances among all evaluated instances, offering an overall measure of the model’s performance:
$Accuracy = \frac{TP + TN}{TP + TN + FN + FP}.$

\subsection{Baseline Methods}
In our evaluation, we compare CSLS with nine state-of-the-art methods.

(1) ChatGPT~\cite{ChatGPT}: The ChatGPT model demonstrates the capabilities of deep learning in code generation and processing, though it is not specifically designed for software vulnerability detection. 
We tested the dateset on chain-of-though based cue words in ChatGPT 3.5 and ChatGPT 4o, repeated the process three times, and averaged the results~\cite{wei2022chain}. 
We provide the prompt we use in the anonymous code package.

(2) Devign~\cite{zhou2019devign}: Devign utilizes a graph-based model with a Gated Graph Recurrent Network (GGN) to represent the graph that combines the Abstract Syntax Tree (AST), Control Flow Graph (CFG), Data Flow Graph (DFG), and code sequence of the input code fragment for vulnerability detection.

(3) ReGVD~\cite{nguyen2022regvd}: By transforming the source code into a graph structure, ReGVD uses the label embedding in GraphCodeBERT to learn the code structure and enhance the vulnerability detection ability of the code model.


(4) CodeBERT~\cite{feng2020codebert}: CodeBERT is a pre-trained model that integrates natural language and programming language representations, supporting a wide range of coding tasks, including code understanding and generation.

(5) CodeT5+~\cite{wang2023codet5+}: CodeT5+ is an encoder-decoder model for code, with flexible component modules that can be combined to handle a variety of downstream code tasks.

(6) UniXcoder~\cite{guo2022unixcoder}:  UniXcoder extends the capabilities of models like CodeBERT by incorporating a deep understanding of code syntax and semantics, thereby enhancing model performance on tasks such as code summarization, translation, and completion. UniXcoder-nine, its latest extension in 2023, further trains UniXcoder-base to develop robust code models.

(7) TRACED~\cite{ding2024traced}: TRACED introduces an execution-aware pre-training strategy for source code by integrating execution traces with source code and executable inputs. This method enhances the model’s ability in tasks like static execution estimation, clone retrieval, and vulnerability detection. 


\subsection{Implementation}
CSLS contains a pre-trained model fine-tuning process and is therefore affected by the learning rate and Batch size. We provide detailed runtime environment and script setup in the open source code repository. Due to page limitations, we do not discuss the impact of different Settings in this category. In general, a batchsize of 12 and a learning rate of 2e-5 is a recommended choice.

All experimental procedures are saved with the random seed used in the existing literature: seeds=123456~\cite{nguyen2022regvd}. All experiments are repeated 3 times to ensure repeatability.

\textbf{Scope:} CSLS currently identifies vulnerabilities at the function level. However, recent works, such as that by Li et al.~\cite{li2021vulnerability}, utilize deep learning explanation tools to report vulnerabilities at the line level. By leveraging our approach's ability to model line-level semantics, we can directly pinpoint indices of line-level vulnerabilities, identifying them as dangerous rows. Nevertheless, comprehensive verification of these indices necessitates substantial expert knowledge or a standardized verification methodology, which we defer to future work.

\begin{figure*}[t]
    \centering
    \begin{subfigure}[t]{0.24\textwidth}
        \centering
        \includegraphics[width=\textwidth]{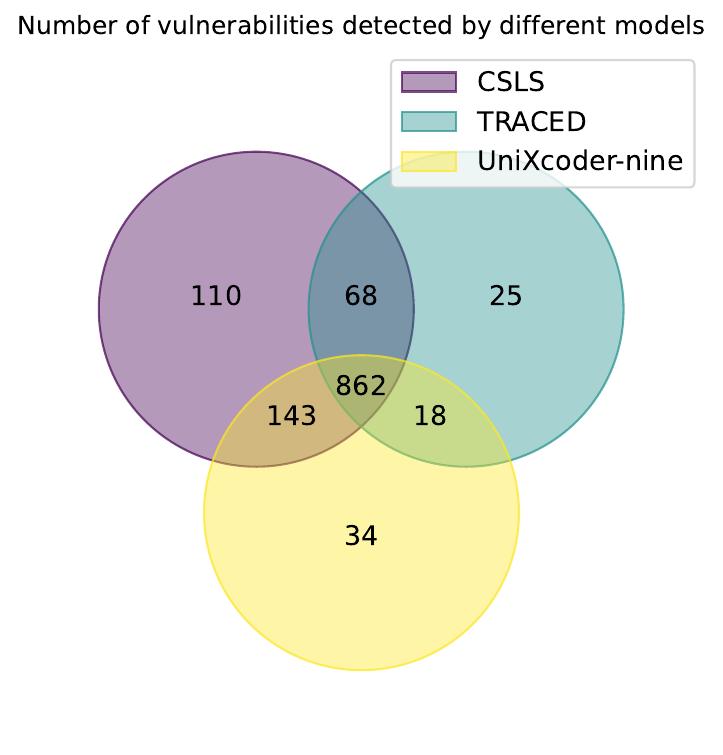}
            \vspace{-2em}
        \caption{Number of vulnerabilities discovered in dataset Devign}
        \label{fig:result_d_venn}
    \end{subfigure}
    \hfill
    \begin{subfigure}[t]{0.24\textwidth}
        \centering
        \includegraphics[width=\textwidth]{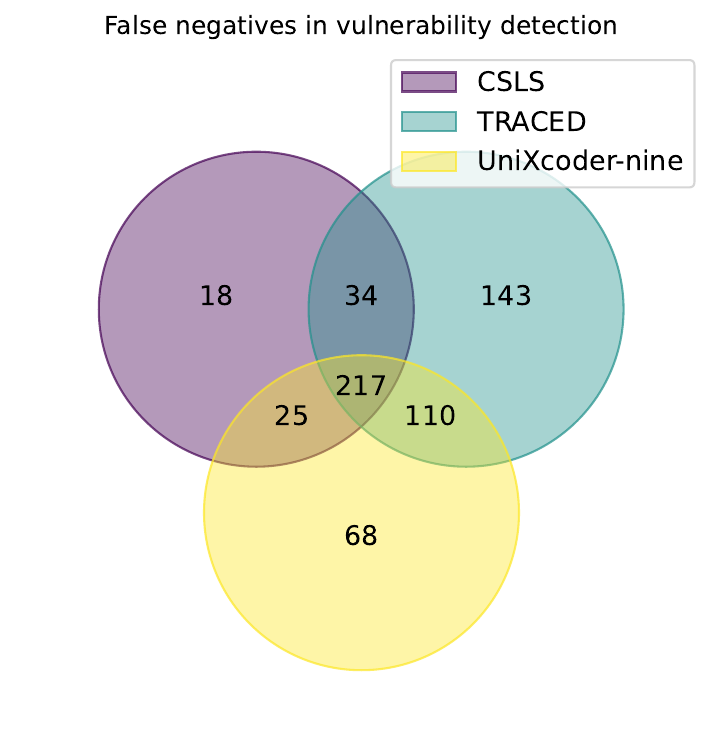}
          \vspace{-2em}
        \caption{Number of missed reports in dataset Devign}
        \label{fig:result_fn_d_venn}
    \end{subfigure}
    \hfill
    \begin{subfigure}[t]{0.24\textwidth}
        \centering
        \includegraphics[width=\textwidth]{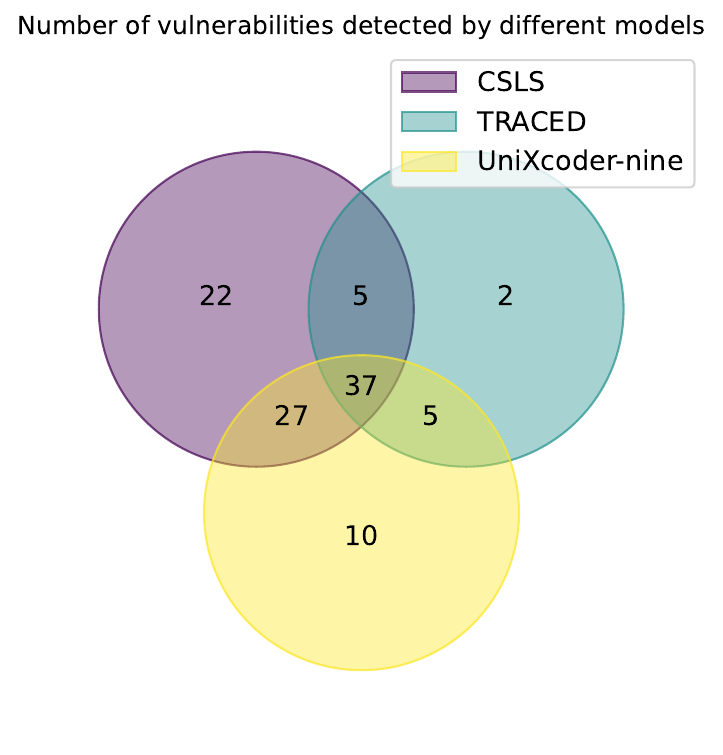}
        \vspace{-2em}
        \caption{Number of vulnerabilities discovered in dataset Reveal}
        \label{fig:ruselt_rv_fn_venn}
    \end{subfigure}
    \hfill
    \begin{subfigure}[t]{0.24\textwidth}
        \centering
        \includegraphics[width=\textwidth]{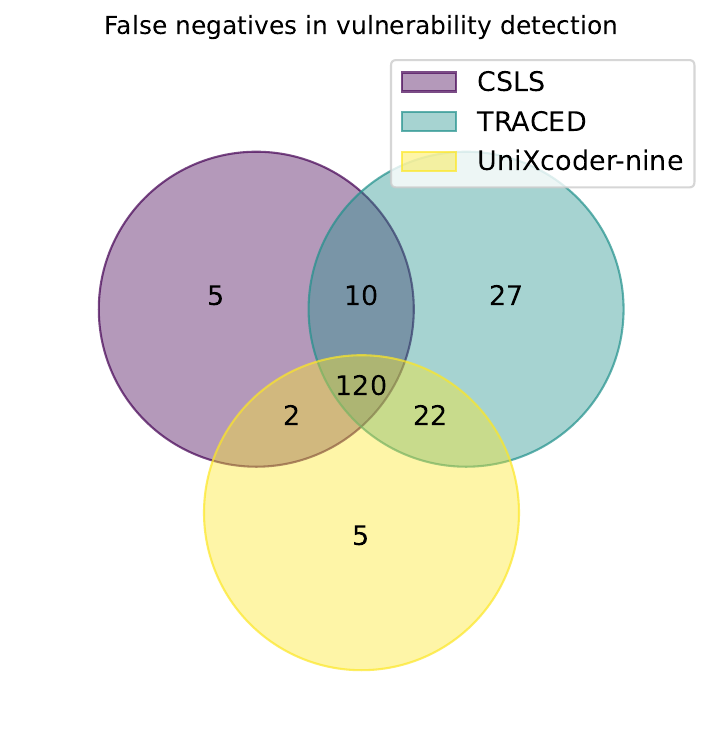}
          \vspace{-2em}
        \caption{Number of missed reports in dataset Reveal}
        \label{fig:ruselt_rv_venn}
    \end{subfigure}
    \caption{Comparison of vulnerability detection performance of different models on two datasets.}
    \label{fig:venn_diagrams}
    \vspace{-1em}
\end{figure*}

\section{EXPERIMENTS}
We aim to answer the following Research Questions (RQs):

\textbf{RQ1:} How effective is CSLS compared with the state-of-the-art baselines on vulnerability detection?

\textbf{RQ2:} What are the effects of the code pre-processing process on vulnerability detection methods based on fine-tuning of pre-trained models?


\textbf{RQ3:} What is the difference in the performance of the model using different pre-trained codes?

\subsection{RQ1. Effectiveness of CSLS }
To answer the first question, we compare CSLS with the seven baseline methods on the two datasets as shown in Table 1. 
We can draw conclusions about the performance of CSLS compared to the baselines across the evaluated datasets. 

\begin{table}[t]
\centering
\caption{Number of tokens for different datasets after preprocessing. T stands for the number of tokens.}
\label{toekn}
\small
\begin{threeparttable}
\begin{tabularx}{\linewidth}{lXXXXX}
\toprule
 Data  & Our(T) & Exist(T) & Ratio &line\_num&T$>$1024 \\ \midrule

Devign & 723.70 & 559.74 & 77\% &113.37&14.64\%\\

Reveal  & 488.40&457.34&94\%&32.07 &8.95\%\\
PrimeVul*&449.12&360.73&80\%&44.94&0\%\\

\bottomrule
\end{tabularx}
\end{threeparttable}
\vspace{-1em}
\end{table}

\begin{table}[t]
\centering
\caption{Comparison results for different models on Devign dataset. (S) stands for preserving structural information. L represents the maximum input length of the model. }
\label{tab:3}
\small
\begin{threeparttable}
\begin{tabularx}{\linewidth}{lXXXXX}
\toprule
 & \multicolumn{5}{c}{\textbf{Devign \cite{zhou2019devign}}} \\ \cline{2-6} 
 Models& L & Acc& Recall & Prec & F1 \\ \midrule

CodeBERT& 400& 63.59 & 41.99 & 66.37 & 51.43 \\

CodeBERT(S) &400& 65.95& 44.46 &70.54&54.54 \\
\hline
CodeT5+-base&512& 65.62 & 55.29 & 64.73 & 59.64 \\ 

CodeT5+-base(S)& 512& 66.43 & 44.38 & 71.77 & 59.64\\ 

TRACED &512& 64.42 & 61.27 & 60.03 & 61.05 \\ 
TRACED(S)& 512& 67.45& 54.26& 68.37& 60.50
\\ 
\hline
UniXcoder-base&1024 & 65.77 & 51.55 & 66.42 & 58.05 \\ 
UniXcoder-base(S)& 1024& 68.85 & 58.64& 68.91 & 63.36
 \\



UniXcoder-nine &1024& 66.98 & 56.33 & 66.63 & 61.05 \\ 
UniXcoder-nine(S) &1024&69.10& 56.73& 70.28 & 62.78
 \\ 
\bottomrule
\end{tabularx}

\end{threeparttable}
\vspace{-2em}
\end{table}

\begin{table}[t]
\centering
\caption{Results of statistical tests for model comparison.}
\label{tab:chi_square}
\footnotesize
\setlength{\tabcolsep}{3pt} 
\begin{threeparttable}
\begin{tabular}{>{\raggedright}p{4cm} >{\centering\arraybackslash}p{2cm} >{\centering\arraybackslash}p{2cm}}
\toprule
Models compared & $\chi^2$ statistic & p-value \\ \midrule
CodeBERT vs. CodeBERT(S) & 697.50 & $1.04 \times 10^{-153}$ \\
Traced vs. Traced(S) & 799.29 & $7.66 \times 10^{-176}$ \\
UniXcoder-b vs. UniXcoder-b(S) & 809.34 & $5.01 \times 10^{-178}$ \\
UniXcoder-n vs. UniXcoder-n(S) & 696.84 & $1.45 \times 10^{-173}$ \\
CSLS vs. UniXcoder-n(S) & 1433.51 & $0$ \\
\bottomrule
\end{tabular}
\end{threeparttable}
\vspace{-2em}
\end{table}

Table 1 presents the performance of ChatGPT on the vulnerability detection task. It is evident that large-scale language models employ aggressive detection logic. Specifically, in ChatGPT 4o, nearly all code snippets were identified as vulnerable, leading to considerably low F1 scores across both datasets.

CSLS demonstrates a marked superiority on both datasets. 
In the Devign dataset, CSLS attains the highest Accuracy of 70.57\%, the highest F1 score of 64.95\% and the highest Precision outperforming all other models.  This improvement can be attributed to the retention of structural information during pre-processing, which enhances the model's understanding of code semantics.
We provide a Venn diagram of the performance of different models for vulnerability detection in Fig. 4, where the false positive rate of each method does not significantly differ according to the metric data; therefore, we pay more attention to the accuracy and false negative rate of the model for vulnerabilities.
As shown in Fig. 4(a), CSLS detected 110 different vulnerabilities, while TRACED detected 68 and UniXcoder-nine detected 25. This indicates that CSLS has a higher detection capability, identifying more vulnerabilities than the other models. In terms of missed reports, CSLS had the fewest false negatives (18), whereas TRACED and UniXcoder nine missed 34 and 143 vulnerabilities, respectively. This demonstrates CSLS's superior accuracy in minimizing missed detection as shown in Fig. 4(b).
This indicates that CSLS has the most balanced performance in correctly identifying vulnerabilities without being skewed towards over-predicting (which would increase recall but decrease precision) or under-predicting (which would do the opposite).

On the Reveal dataset, since the proportion of negative samples in this dataset is 90, the model with strong fitting performance generally exceeds 90\% on ACC.
In this dataset, people are generally interested in the ability of the model to find positive samples (vulnerability).
For CSLS, both Recall and F1 metrics maintain the level of optimal level.
As show in Fig. 4(c), CSLS detected 22 vulnerabilities, outperforming TRACED (7) and UniXcoder-nine (2). This again shows CSLS's higher effectiveness in detecting vulnerabilities. In Fig. 4(b), CSLS had 10 missed reports, which is significantly lower than TRACED (27) and UniXcoder nine (5), indicating a balanced performance in both detecting vulnerabilities and minimizing false negatives.
These figures not only show that CSLS maintains its high performance in different testing conditions but also that it consistently understands and predicts code vulnerabilities with high precision and recall. 
%


\begin{table}[t]
\centering
\caption{Comparison results for different models on Reveal dataset. (S) stands for preserving structural information.}
\label{tab:reveal}
\small
\begin{threeparttable}
\begin{tabularx}{\linewidth}{lXXXX}
\toprule
 & \multicolumn{4}{c}{\textbf{Reveal \cite{chakraborty2021deep}}} \\ \cline{2-5} 
 Models & \textbf{Acc} & \textbf{Recall} & \textbf{Prec} & \textbf{F1} \\ \midrule

CodeBERT & 90.10 & 28.50& 51.18 & 36.61 \\

CodeBERT(S) & 89.18& 29.80 &44.15& 35.60 \\\hline
CodeT5+-base&  90.94 & 31.14& 59.16& 40.80 \\ 

CodeT5+-base(S)& 91.20 & 29.38 & 63.20& 40.11\\ 

TRACED & 90.12& 24.14  & 67.07 & 35.48 \\ 
TRACED(S) & 91.95& 24.12& 84.61& 37.54
\\ \hline
UniXcoder-base & 90.50 & 39.91 & 53.52 & 45.72\\ 
UniXcoder-base(S) & 90.50 & 42.98& 53.26&47.57
 \\

UniXcoder-nine & 90.14& 31.57& 51.42 & 39.13 \\ 
UniXcoder-nine(S) &91.33& 27.63& 66.31 & 39.00
 \\ 

\bottomrule
\end{tabularx}

\end{threeparttable}
\vspace{-2em}
\end{table}

\subsection{RQ2: Effect of different code pre-processing procedures on detection performance}
To illustrate the impact of the code pre-processing process on the vulnerability detection task, we conduct experiments on several models. Since we propose to keep line characters and Spaces in code snippets, this increases the number of tokens per sample. This can lead to bad results in cases where the input length of the pre-trained model is limited.
Therefore, we analyze the token length after code-word segmentation.

As can be seen from Table \ref{toekn}, the situation of the number of tokens after PBE word segmentation after different preprocessing methods have been adopted for different data sets.
The data shows that the Devign dataset is severely affected, with the number of tokens reduced by 23\% after processing by the currently commonly used pre-processing method (Exist). At the same time, the Devign dataset is also far more than other datasets in the number of lines of code.
The code snippet in Reveal is much less affected. There is possible reasons for this. First, the complexity of the code function of the projects in Reveal is lower than that of Devign, and the average number of tokens is only half of that of Devign. At the same time, the proportion of code functions with more than 1024 tokens in Devign is more than 14.64, which is much higher than that of Reveal data. 
In order to better verify the effectiveness, we sampled a new vulnerability detection benchmark PrimeVul~\cite{ding2024vulnerability}, which has more diverse vulnerability types and more extreme data imbalance. 
There are some samples in this dataset that are far longer than the input length of the baseline model, so we remove the samples that tokens are beyond 1024 to obtain PrimeVul*.
\begin{table}
\centering
\caption{Comparison results for different models on PrimeVul dataset. (S) stands for preserving structural information. }
\label{tab:PrimeVul}
\small
\begin{threeparttable}
\begin{tabularx}{\linewidth}{lXXXX}
\toprule
 & \multicolumn{4}{c}{\textbf{PrimeVul \cite{ding2024vulnerability}}} \\ \cline{2-5} 
 Models & Acc & Recall & Prec & F1 \\ \midrule
Unixcoder-BASE & 55.76 & 67.17 & 55.12 & 60.54 \\ 
Unixcoder-BASE (S) & 56.68 & 70.21 & 55.66 & 62.09 \\ 
Unixcoder-nine & 56.22 & 49.54 & 57.80 & 53.35 \\ 
Unixcoder-nine (S) & 56.83 & 68.96 & 55.91 & 61.76 \\ 
Our (Unixcoder-nine) & 56.98 & 76.59 & 55.38 & 64.28 \\ \bottomrule
\end{tabularx}
\end{threeparttable}
\vspace{-1em}
\end{table}

Table ~\ref{tab:3} shows the vulnerability detection performance on the dataset Devign under different data pre-processing processes.
The experimental results show that all the models have a great degree of performance improvement compared with the indicators reported in the literature.
For example, CodeBERT, whose best known performance reported in the literature was only 63\%@ACC, went up to 65.95\%@ACC using the new code pre-processing pipeline without making any model modifications. 
At the same time, it is clear that models with input sequence lengths of 1024 get a bigger boost because they can see more structural information while preserving the semantics of the code. Models with lower input lengths often have to trim their code. 
%
Table ~\ref{tab:chi_square} presents the statistical results of vulnerability detection based on different preprocessing methods. The data show that different preprocessing methods make significant differences in model checking results.
%

Table ~\ref{tab:reveal} shows the situation on Reveal, where there is a significant difference between Reveal and the dataset Devign. 
Firstly, the two processing methods have less impact on this dataset than Devign. Secondly, there are far more non-vulnerability data than vulnerability data in this data, which increases the difficulty of vulnerability detection.
The performance of CodeBERT decreases after our way of pre-processing. This is due to the limited input, the model learns fewer tokens after introducing structural information, and the unbalanced samples lead to a decrease in the modeling ability of the model.
As the input length of the model increases, the other models get better, and the performance of different models on this dataset improves. This may be due to the fact that this dataset has much less structural information than Devign (only 6\%).
As shown in Table ~\ref{tab:PrimeVul}, in the PrimeVul dataset, the model performs poorly, but the trend is consistent as other datasets.



\subsection{RQ3. Effects of using different code models in CSLS}
\begin{table}[t]
\centering
\caption{Effects of different code models. CSLS(X) means that the global semantic model and the row-level semantic model use the model.}
\label{DIFFMODEL}
\small
\begin{threeparttable}
\begin{tabularx}{\linewidth}{lXXXX}
\toprule
 & \multicolumn{4}{c}{\textbf{Devign \cite{zhou2019devign}}} \\ \cline{2-5} 
 Models & Acc& Recall & Prec & F1 \\ \midrule

CodeBERT(CB) & 63.59 & 41.99 & 66.37 & 51.43 \\
CSLS(CB)& 65.84 & 44.39& 70.32& 54.42
 \\
 TRACED(T) & 64.42 & 61.27 & 60.03 & 61.05 \\ 
 CSLS(T)& 67.60 & 57.37 & 67.28 & 61.93 \\ 
UniXcoder-base(Un-b)& 65.77 & 51.55 & 66.42 & 58.05 \\ 

CSLS(Un-b) & 69.43 & 58.08 &70.23& 63.58 \\ 
UniXcoder-nine(Un-n)&66.98 & 56.33 & 66.63 & 61.05 \\ 
CSLS(Un-n) & 70.57 & 59.36 & 71.70 & 64.95 \\ \bottomrule
\end{tabularx}
\end{threeparttable}
\vspace{-1em}
\end{table}
In this section, we investigate the effects of using different code models in the Code-Semantic Learning System (CSLS). The experiment involves two code models, allowing for various combinations. The results, as summarized in Table V, demonstrate that our method provides enhancements regardless of the base model used.

Table ~\ref{DIFFMODEL} presents a comparison of different models on the Devign dataset. The metrics considered include Accuracy (Acc), Recall, Precision (Prec), and F1 score. Here are the detailed observations from the results: Table 7 shows that combining different code models leads to improved performance, with UniXcoder models showing significant gains. The CSLS(Un-N) achieves the highest metrics, including an accuracy of 70.57\%, recall of 59.36\%, precision of 71.70\%, and an F1 score of 64.95\%. These results indicate that leveraging different model configurations can significantly enhance vulnerability detection performance.

From these results, it is evident that using different combinations of code models in CSLS leads to consistent improvements across all evaluated metrics. Notably, the combinations involving UniXcoder exhibit significant enhancements, with the CSLS(Un-N) standing out as the most effective. These experimental results verify that our method can be effectively extended based on different code models for different application scenarios.


\subsection{Hyperparameter Experiments}
CSLS has two hyperparameters: the number of lines \(k\) and the default number of tokens per line \(p\). Given the strong domain-specific nature of the vulnerability detection task, it is advisable to choose the settings of these hyperparameters based on the scenario. For the experiments in this paper, we selected the hyperparameter settings according to the code style of the projects included in the Devign dataset.

As shown in Table ~\ref{pandk}, we experimented with different combinations of \(p\) and \(k\). Specifically, we tested \(p\) values of 10 and 20, and \(k\) values of 70, 100, and 120. The results indicate that the best performance in terms of accuracy (70.57) and F1-score (64.95) was achieved with \(p = 20\) and \(k = 100\). We did not explore other combinations of \(p\) and \(k\) as the average number of lines in the code snippets of this dataset is 117, with an average of 15 tokens per line.

\begin{table}[t]
\centering
\caption{Comparison results for different Hyperparameter. }
\label{pandk}
\small
\begin{threeparttable}
\begin{tabularx}{\linewidth}{lXXXXX}
\toprule
\multicolumn{1}{c}{} & \multicolumn{1}{l}{\textbf{Hyper-parameter}} & \multicolumn{4}{c}{\textbf{Devign \cite{zhou2019devign}}} \\ \cline{2-6} 
 P &K & Acc& Recall & Prec & F1 \\ \midrule

20&70& 69.32 & \textbf{63.10}&67.86& 65.40 \\ 
20 &100 & \textbf{70.57} & 59.36 & 71.70 & 64.95 \\ 
20&120&69.21& 51.47& \textbf{73.57} & 61.07\\ 
10&70& 69.61 & 61.59 & 68.95 & \textbf{65.06} \\ 
10&100 &69.83 &60.23& 69.93& 64.72\\
10&120& 69.10 & 59.92 & 68.80& 64.05 \\

\bottomrule
\end{tabularx}
\end{threeparttable}
\vspace{-1em}
\end{table}

\section{Discussion}

\subsection{Influence of structural information on positive and negative samples }
A question of interest is how structural information plays a role in vulnerability detection. 
An experiment was used to evaluate the impact of different preprocessing flows on positive (vulnerable) and negative (non-vulnerable) samples in the dataset.  Specifically, the dataset is pre-processed in two distinct ways, and then the UniXcoder model is fine-tuned for four epochs. 
The global CLS semantic features are subsequently reduced to a two-dimensional classification plane. In the figures provided, label 0 represents vulnerable samples, and label 1 represents non-vulnerable samples. Fig. 5(a) and Fig. 5(b) illustrate the results of the two pre-processing methods.
\begin{figure}[t]
    \centering
    \begin{subfigure}[t]{0.22\textwidth}
        \centering
        \includegraphics[width=\textwidth]{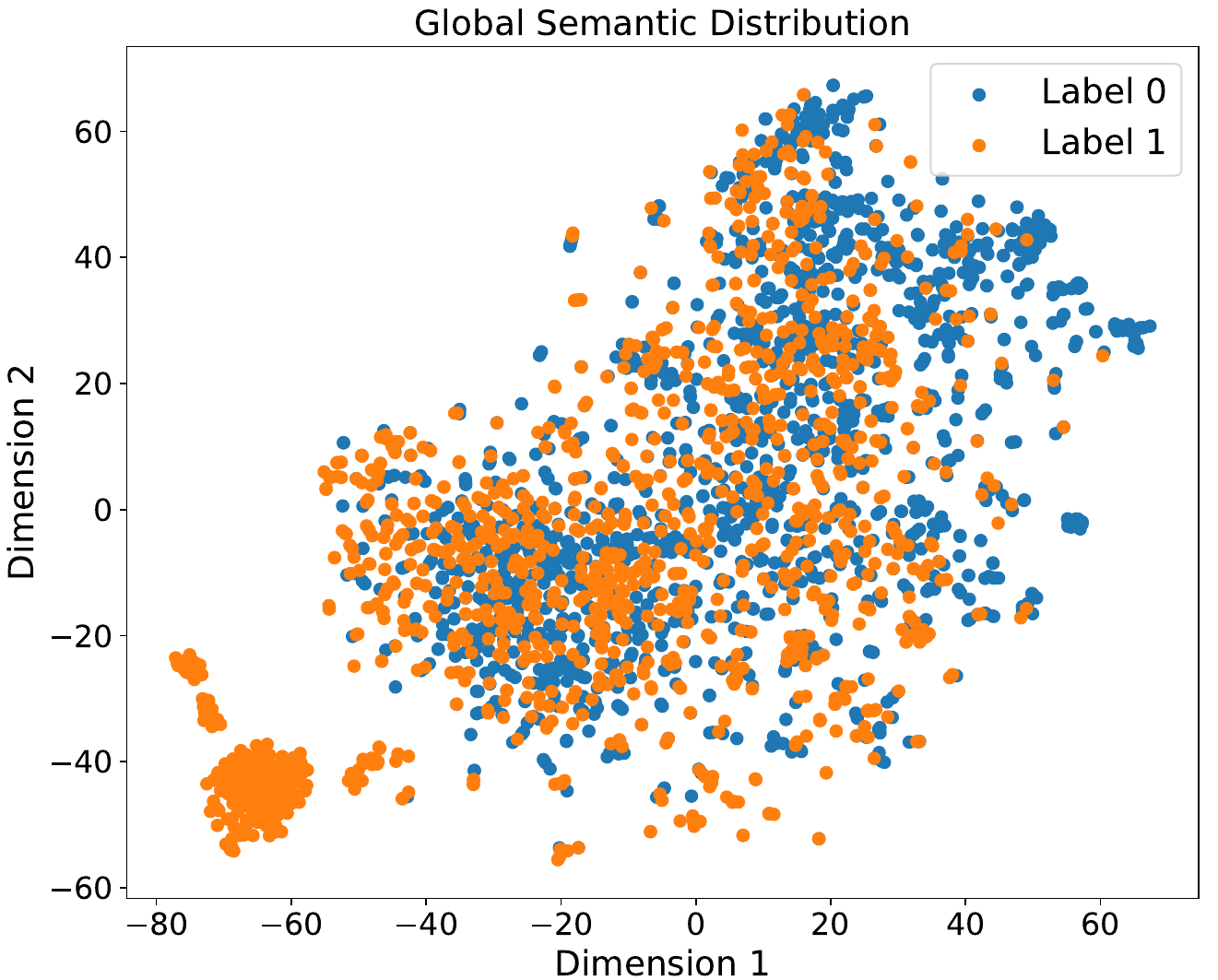}
        \caption{Pre-processing that preserves the code structure}
        \label{fig:result_d_venn}
    \end{subfigure}
    \hfill
    \begin{subfigure}[t]{0.24\textwidth}
        \centering
        \includegraphics[width=\textwidth]{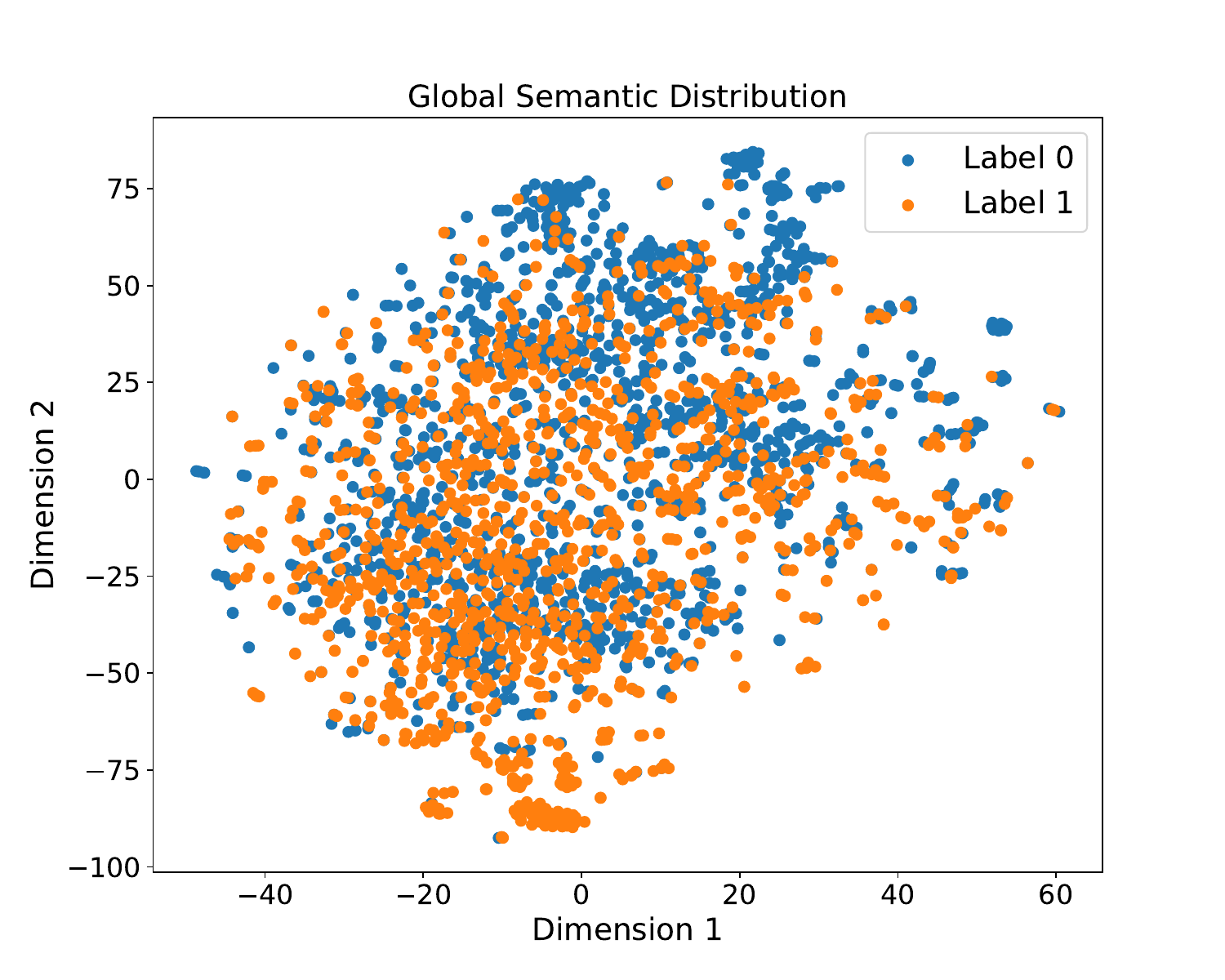}
        \caption{Pre-processing that does not preserve the code structure}
        \label{fig:result_fn_d_venn}
    \end{subfigure}
  
    \caption{Comparison of vulnerability detection performance of different prep-rocessing process.}
    \label{fig:venn_diagrams}
    \vspace{-2em}
\end{figure}
Fig. 5(a), the pre-processing method preserves the structural information of the dataset Devign.
Conversely, in Fig. 5(b), the pre-processing method does not preserve the structural information. 
This is evident from the distinct clustering of the positive (blue) and negative (orange) samples.
The separation between the two clusters suggests that the preserved structural information facilitates better differentiation between vulnerable and non-vulnerable samples. 
As a result, the classifier can more easily identify and segregate these samples based on their inherent characteristics.

%

As shown in Fig. 5(a), the non-vulnerable samples (located in the lower left) exhibit clear clustering. 
This suggests that such a pre-processing method is more beneficial for the model in classifying positive samples.

\section{THREATS TO VALIDITY}
Threats: We used multiple code models as well as a transformer model for the vulnerability detection task, which may increase the model parameters (297.74M).
This can lead to higher training and deployment costs. However, due to the scarcity of vulnerability data, researchers generally can only use smaller code models (110M-770M) to complete fine-tuning, so we believe that this overhead is in an acceptable range. Inside. At the same time, the batch design with line-level semantic awareness does not show a multiple increase in memory footprint.

Internal Validity: During validation on different datasets, we found that CSLS is affected by code style. Code with a minimalist style is likely to be less affected by the approach as they almost have similar code structure. The vulnerability detection of code with complex structure is more beneficial than the structure-based detection model. At the same time, even if we maintain a good expectation for the input length of the pre-trained code model, it is undeniable that our pre-training method increases the input length, which may lead to the performance of CSLS being affected in some restricted scenarios.

\section{Related work}

\subsection{Traditional Vulnerability Detection}
Over the years, numerous methods for vulnerability detection have been developed. Initially, research in this area predominantly focused on identifying vulnerabilities through manually customized rules~\cite{Checkmarx,Flawfinder}.
Static analysis tools rely on manual rules and precise specification of code behavior, which are difficult to obtain automatically. While these heuristic approaches offered solutions for vulnerability detection, they necessitated extensive manual analysis and the formulation of defect patterns. 
 %

\subsection{Deep Neural Network for Vulnerability Detection}
To perceive code text nonlinear characteristics, recent research has turned to the model based on neural network, hole features extracted from code snippets ~\cite{dam2017automatic, russell2018automated}.
Existing deep learning-based vulnerability detection models are predominantly divided into two categories: token-based and graph-based models.

Token-based models treat code as a linear sequence and use neural networks (e.g., LSTM or Transformer) to learn vulnerability features from known cases, aiming to identify vulnerability features~\cite{russell2018automated,li2018vuldeepecker,cheng2021deepwukong}. 
%
%
Concurrently, Li et al.\cite{li2018vuldeepecker} employed BiLSTM~\cite{schuster1997bidirectional} to encode a segmented version of input code, known as 'code gadgets,' centered on key markers, especially library/API function calls.
However, these token-based models often overlook the complexity of the source code structure, potentially leading to inaccurate detection.
%

In parallel, another research direction explores the potential of graph-based methods for vulnerability detection~\cite{li2021vulnerability,chakraborty2021deep,zheng2021vu1spg,wu2022vulcnn,cao2022mvd}. 
For example, DeepWukong~\cite{cheng2021deepwukong} uses GNNs for feature learning, focusing on compressing code fragments into a dense, low-dimensional vector space to enhance the detection of various vulnerability types.
%
%
Graph-based detection models learn code structure through various graph representations, utilizing neural networks for vulnerability detection~\cite{wu2022vulcnn,cao2022mvd}. 
For instance, Zhou et al.\cite{zhou2019devign} used a gated graph recurrent network\cite{li2016gated} to extract structural details from triadic graph representations—AST, CFG, and DFG.
Chakraborty et al.\cite{chakraborty2021deep} introduced REVEAL, an innovative approach that combines a gated graph neural network, re-sampling techniques\cite{chawla2002smote}, and triplet loss~\cite{mao2019metric}.
%
%

\subsection{Pre-Trained Models for Vulnerability Detection}

Inspired by the success of pre-trained models in natural language processing (NLP), recent research has increasingly focused on leveraging these models to enhance code vulnerability detection accuracy~\cite{feng2020codebert,kanade2020learning,niu2022spt,lin2021traceability,bai2021syntax,guo2022unixcoder,lachaux2021dobf,wang2024m2cvd}. 

The core concept behind these works is to use a model pre-trained on a large corpus of source code data, followed by specialized fine-tuning for specific tasks~\cite{kanade2020learning}.
For instance, Feng et al.\cite{feng2020codebert} proposed CodeBERT, designed specifically for understanding and generating source code, combining the processing capabilities of both natural and programming languages.
Similarly, CuBERT employs masked language modeling with sentence prediction for code representation\cite{kanade2020learning}. 
Additionally, some pre-trained models incorporate structural information of code fragments during the initial training phase~\cite{niu2022spt,lin2021traceability}. 
For example, Guo et al.'s GraphCodeBERT~\cite{guo2020graphcodebert} leverages graph structures to infer data flow in code fragments.
%
%
The objective is specifically tailored to address the structural dimension of programming languages.
In comparative evaluations, CodeBERT is positioned as a baseline standard for various code-related tasks, including code clone detection and code translation. 
UniXcoder, a unified cross-modal pre-trained programming language model, also serves as a baseline method, trained on extensive code and natural language data~\cite{guo2022unixcoder}.

%
Hanif et al. proposed VulBERTa, which pretrains a RoBERTa model with a custom tokenization pipeline for real-world C/C++ projects~\cite{hanif2022vulberta}.
Nguyen et al. introduced ReGVD, combining graph structure and pre-trained models to address source code vulnerability detection~\cite{nguyen2022regvd}. 
Zhang et al. decomposed the code segment Control Flow Graph (CFG) into multiple execution paths and used the pre-trained model CodeBERT for vulnerability detection~\cite{zhang2023vulnerability}. 
and Thapa et al.~\cite{thapa2022transformer} explored the performance of fine-tuned language models for multi-class classification of similar types of vulnerabilities.

%

\section{CONCLUSION}
In this paper, we introduced the Code Structure-Aware Network through Line-level Semantic Learning (CSLS) to enhance code vulnerability detection. 
Our approach involves a refined code text processing workflow that retains structural elements, such as newlines and indentation, before modeling. This allows our proposed CSLS architecture to effectively capture and utilize line-level structural and semantic information. 
The CSLS architecture integrates four components: code preprocessing, global semantics-aware, line semantic-aware, and line semantic structure-aware.
The core idea of CSLS architecture is to realize nonlinear structure modeling of row-level semantics in the row-level semantic space to achieve high-precision vulnerability detection.

Future work will focus on further refining the model architecture and exploring its applicability to other programming languages and broader categories of software vulnerabilities. Additionally, we aim to investigate the impact of different pre-processing techniques and model configurations to further enhance the detection capabilities of CSLS.

\bibliographystyle{IEEEtran}
\bibliography{sample-base}


\end{document}